\documentclass[aps,prb,twocolumn]{revtex4} %,preprint

\usepackage{graphicx}% Include figure files
\usepackage{dcolumn}% Align table columns on decimal point
\usepackage{bm}% bold math

\newcommand{\comment}[1]{}
\begin{document}
%\preprint{}   % Preprint number in upper right corner
\renewcommand{\theequation}{\arabic{section}.\arabic{equation}}

\title{New Theory for Cooper Pair Formation and Superconductivity}

%\author{}
%\thanks{}

\author{Phil Attard}
\affiliation{
{\tt phil.attard1@gmail.com}}
%\\ 23 Mar., 2022}
%\noindent {\tt  Projects/QSM22/Paper2/nts.v3.tex
%\\ Original published as
%arXiv:2203.12103v1} %\newline
%{\tt 6--12 Feb., 2022}
%\affiliation{\protect\texttt{phil.attard1@gmail.com}}

%\date{\today. Begun  27Feb22, phil.attard1@gmail.com
% version 1, nts.v1.tex differs significantly; ideal results same
%\\ notes in My Documents/Projects/QSM22/supercon.tex}

\begin{abstract}
A new theory for Cooper pair formation and superconductivity is derived
from  quantum statistical mechanics.
It is shown that zero momentum Cooper pairs
have non-local permutations and behave as effective bosons
with an internal weight close to unity
when bound by a primary minimum
in the potential of mean force.
For a short-ranged, shallow, and highly curved  minimum
there is no thermodynamic barrier to condensation.
The size of the condensing Cooper pairs found here
is orders of magnitude smaller than those found in BCS theory.
The new statistical theory is applicable
to high temperature superconductors.
%Computational results for Lennard-Jones $^3$He
%show that the entropy barrier to condensation
%is extrapolated to vanish at about  0\,K.
\end{abstract}

%\keywords{}

\maketitle

%\newpage
%%%%%%%%%%%%%%%%%%%%%%%%%%%%%%%%%%%%%%%%%%%%%%%%%%%%%%%%%%%%%%%%%%%%%%%%%%
%
\section{Introduction}
\setcounter{equation}{0} \setcounter{subsubsection}{0}
%\renewcommand{\theequation}{\arabic{section}.\arabic{equation}}
%\renewcommand{\theequation}{\Alph{section}.\arabic{equation}}
%
%%%%%%%%%%%%%%%%%%%%%%%%%%%%%%%%%%%%%%%%%%%%%%%%%%%%%%%%%%%%%%%%%%%%%%%%%%

Superconductivity has long been described
by the  Bardeen-Cooper-Schrieffer (BCS) theory.\cite{BCS57}
This invokes the general notion of Cooper pairs
(electrons with equal and opposite momentum and spin,
bound by an attractive potential),\cite{Cooper56}
together with the specific proposal
that the attraction is due to
the dynamic interaction of the electron pair with
the quantized vibrations of the solid lattice,
which explains the dependence of the transition temperature
on the isotopic masses of the solid.\cite{Maxwell50,Reynolds50}
The discovery of high temperature superconductors,\cite{Bednorz86,Wu87}
which show no dependence on the isotopic mass,
rules out the specific phonon exchange mechanism in these cases,
although it leaves open the possibility for Cooper pairing
if another mechanism for a binding potential could be found.
Despite many proposals
\cite{Anderson87,Bickers87,Inui88,Gros88,Kotliar88,Mann11,Monthoux91}
no consensus for such a potential has emerged.

I also have proposed a specific mechanism
for an attraction between electrons,\cite{Attard22b}
namely the monotonic-oscillatory transition
that generically occurs at high coupling in charge systems.
The transition appears realistic in the relevant regime
for high temperature superconductors.\cite{Attard22b}
Unlike the generic BCS theory,
the attraction is due to the pair potential of mean force
rather than to the pair potential energy,
and it also appears on much smaller length scales.
For the proposed  monotonic-oscillatory transition
to make sense as the mechanism for Cooper pair formation
in high temperature superconductors,
it is necessary to show that
it is an attractive potential of mean force,
rather than an attractive interaction potential,
that drives Cooper pair formation.

The difference between the interaction potential
and the potential of mean force
is akin to the difference between quantum mechanics
and quantum statistical mechanics.
This paper develops a new theory of superconductivity
based on my formulation of quantum statistical mechanics
in classical phase space,\cite{Attard18,Attard21}
and is analogous to my recent theory
for superfluidity.\cite{Attard21,Attard22a}
The bound Cooper pairs invoked here appear qualitatively different
to those in BCS theory,
at least for their size and statistical binding mechanism,
which suggests that the present statistical thermodynamic theory may be
the one applicable to high temperature superconductors.

%\newpage
%%%%%%%%%%%%%%%%%%%%%%%%%%%%%%%%%%%%%%%%%%%%%%%%%%%%%%%%%%%%%%%%%%%%%%%%%%
%
\section{General Formulation and Analysis}
\setcounter{equation}{0} \setcounter{subsubsection}{0}
%\renewcommand{\theequation}{\arabic{section}.\arabic{equation}}
%\renewcommand{\theequation}{\Alph{section}.\arabic{equation}}
%
%%%%%%%%%%%%%%%%%%%%%%%%%%%%%%%%%%%%%%%%%%%%%%%%%%%%%%%%%%%%%%%%%%%%%%%%%%

%%%%%%%%%%%%%%%%%%%%%%%%%%%%%%%%%%%%%%%
\subsection{Symmetrization Function For Particles with Spin}
%\label{Sec:SymFn1}

\subsubsection{Symmetrized Wavefunction}

Consider a system of $N$ particles.
The set of commuting dynamical variables for one particle $j$
may be taken to be
${\bf x}_j = \{ {\bf q}_j,\sigma_j\}$,
where ${\bf q}_j = \{q_{jx},q_{jy},q_{jz}\}$
is the position of particle $j$,
and $\sigma_j \in \{-S,-S+1,\ldots,S\}$
is the $z$-component of its spin
(see Messiah section~14.1,
or Merzbacher section~20.5).\cite{Messiah61,Merzbacher70}
For  electrons, $S=1/2$ and $\sigma_j = \pm 1/2$.
Note that here $\sigma$
is \emph{not} a spin operator or a Pauli spin matrix.
Label the $2S+1$ spin eigenstates of particle $j$ by
$s_j \in \{-S,-S+1,\ldots,S\}$,
and the spin basis function by $\alpha_{s_j}(\sigma_j) =
\delta_{s_j,\sigma_j}$.
Note that this is \emph{not} a spinor.
For $N$ particles,
${\bm \sigma} \equiv \{ \sigma_1, \sigma_2, \ldots, \sigma_N \}$,
and similarly for ${\bf s}$ and ${\bf q}$,
and the basis functions for spin space are
$\alpha_{\bf s}({\bm \sigma}) =
\delta_{{\bf s},{\bm \sigma}}
= \prod_{j=1}^N \delta_{s_j,\sigma_j}$.

Because the spin basis functions are Kronecker deltas,
it is easy to show that when symmetrizing
the wave functions, only permutations amongst particles
with the same spin give a non-zero result.
In other words, spin is one characteristic
that identifies identical particles.
This is the reason why two electrons with different spin can occupy
the same single particle state.
I now demonstrate this explicitly.

Let the number of particles
with spin $s$ be  $N_s$, and $N =\sum_s N_s$.
For reasons that will become clear shortly I shall deal with two cases
simultaneously.
The most general case allows permutations $\hat{\mathrm P}$
amongst all $N$ particles irrespective of spin.
There are $M = N!$ such permutations.
The more specialized case only allows permutations amongst particles
with the same spin $\hat{\mathrm P} = \prod_s \hat{\mathrm P}_s$,
the factors of which commute.
There are $M = \prod_s N_s!$ such permutations.

An unsymmetrized  wave function $ \psi({\bf x})$
in general has symmetrized form
(see \S6.4.1 of Ref.~\onlinecite{Attard21})
\begin{equation}
\psi^\pm({\bf x})
\equiv
\frac{1}{\sqrt{\chi^\pm M}}
\sum_{\hat{\mathrm P}} (\pm 1)^{p}
\psi(\hat{\mathrm P}{\bf x}) .
\end{equation}
Normalization gives the symmetrization or overlap factor as
\begin{equation}
\chi^\pm \equiv
\sum_{\hat{\mathrm P}} (\pm 1)^{p}
\langle  \psi(\hat{\mathrm P}{\bf x}) | \psi({\bf x}) \rangle .
\end{equation}
The upper sign is for bosons and the lower sign is for fermions.
Henceforth I  consider only the latter.

Using momentum eigenfunctions with discrete momenta,
the single particle wave function
for fermion $j$ is
\begin{equation}
\Phi_{{\bf p}_j,s_j}({\bf q}_j,\sigma_j)
=
\frac{1}{V^{1/2}} e^{-{\bf p}_j \cdot {\bf q}_j/\mathrm{i}\hbar}
\delta_{s_j,\sigma_j} ,
\end{equation}
with ${\bf p}_j = {\bf n}_j \Delta_p$,
${\bf n}_j$ being a three-dimensional integer,
and $\Delta_p = 2\pi \hbar/L$ being the spacing between momentum states,
with $V=L^3$ being the volume of the cube
to which the particles are confined.
\cite{Messiah61,Merzbacher70}

The symmetrized full system wave function is
\begin{eqnarray}
\Phi^-_{{\bf p},{\bf s}}({\bf q},{\bm \sigma})
& = &
\frac{V^{-N/2}}{\sqrt{\chi^-_{{\bf p},{\bf s}} M}}
\sum_{\hat{\mathrm P}}
(- 1)^{p}
\prod_{j=1}^N
e^{-{\bf p}_{j}' \cdot {\bf q}_j/\mathrm{i}\hbar}
\delta_{s_{j}',\sigma_j}
\nonumber \\ & = &
\frac{V^{-N/2}}{\sqrt{\chi^-_{{\bf p},{\bf s}} M}}
\sum_{\hat{\mathrm P}}
(- 1)^{p}
e^{-{\bf p}' \cdot {\bf q}/\mathrm{i}\hbar}
\delta_{{\bf s}',{\bm \sigma}} .
\end{eqnarray}
The prime signifies the permuted list,
${\bf p}'_j = \{ \hat{\mathrm P} {\bf p} \}_{ j}  $,
and identically for the spin.

The Kronecker-delta $\delta_{{\bf s}',{\bm \sigma}}$
that appears here indicates that only permutations
amongst particles with the same spin need be considered.
Henceforth (until the introduction of pairs)
I take $\hat{\mathrm P} = \prod_s \hat{\mathrm P}_s$
and  $M = \prod_s N_s!$.

%%%%%%%%%%%%%%%%%%%%%%%%%%%%%%%%%%%%%%%%%
\subsubsection{Grand Partition Function}

The grand partition function for fermions is\cite{Attard18,Attard21}
\begin{eqnarray}
\lefteqn{
\Xi^-(z,V,T)
}  \\
& = &
\mbox{TR}' \; e^{-\beta \hat{\cal H}}
\nonumber \\ & = &
\sum_{N} \frac{z^N}{M}
\sum_{\bf p} \sum_{\bf s} \chi^-_{{\bf p},{\bf s}}
\left\langle \Phi^-_{{\bf p},{\bf s}} %({\bf q},{\bm \sigma})
\left| e^{-\beta \hat{\cal H}} \right|
\Phi^-_{{\bf p},{\bf s}} \right\rangle.
\nonumber
\end{eqnarray}
The symmetrization factor counts each state with the correct weight,
which is equivalent to ensuring that each unique allowed state is
counted once only with unit weight.\cite{Attard18,Attard21}
Inserting the above definitions gives
\begin{eqnarray} \label{Eq:Xi-}
\lefteqn{
\Xi^-(z,V,T)
} \nonumber \\
& = &
\sum_{N} \frac{z^N}{\prod_s N_s!}
\sum_{\hat{\mathrm P}} (-1)^p \sum_{\bf p} \sum_{\bf s}
\left\langle
%\hat{\mathrm P} \{{\bf p},{\bf s}\}
\Phi_{\hat{\mathrm P} {\bf p},\hat{\mathrm P} {\bf s} }
\left| e^{-\beta \hat{\cal H}} \right|
\Phi_{ {\bf p},{\bf s} }
%\{{\bf p},{\bf s}\}
\right\rangle
\nonumber \\ & \approx &
\sum_{N} \frac{z^N}{\prod_s N_s!V^{N}}
\sum_{\hat{\mathrm P}} (-1)^p
\sum_{{\bf p},{\bf s}}
\sum_{\bm \sigma}
\nonumber \\ && \mbox{ } \times
\int \mathrm{d}{\bf q}\; e^{-\beta {\cal H}({\bf q},{\bf p})}
e^{{\bf p}' \cdot {\bf q}/\mathrm{i}\hbar}
%\delta_{{\bf s}',{\bm \sigma}} \;
e^{-{\bf p} \cdot {\bf q}/\mathrm{i}\hbar}
\delta_{{\bf s},{\bm \sigma}}
\nonumber \\ & = &
\sum_{N} \frac{z^N}{\prod_s N_s! V^{N}}
\sum_{\hat{\mathrm P}} (-1)^p
\sum_{{\bf p},{\bf s}}
\nonumber \\ && \mbox{ } \times
\int \mathrm{d}{\bf q}\; e^{-\beta {\cal H}({\bf q},{\bf p})}
e^{-[{\bf p}-{\bf p}'] \cdot {\bf q}/\mathrm{i}\hbar}
\nonumber \\ & = &
\sum_{N} \frac{z^N}{\prod_s N_s!V^{N}}
\sum_{{\bf p},{\bf s}} \int \mathrm{d}{\bf q}\;
e^{-\beta {\cal H}({\bf q},{\bf p})}
\eta^-({\bf q},{\bf p},{\bf s}) .
\end{eqnarray}
The second equality neglects the commutation function.\cite{Attard18}
This is a short-ranged function and the approximation is valid
when the system is dominated by long-ranged effects,
which appears to be the case for Bose-Einstein condensation.
The utility of this approximation has been demonstrated
for superfluidity.\cite{Attard21,Attard22a}
Here ${\cal H}({\bf q},{\bf p})
= {\cal K}({\bf p})+U({\bf q}) $ is the Hamiltonian function
of classical phase space.
The present analysis takes the Hamiltonian operator
to be independent of spin.

The symmetrization function is
\begin{eqnarray}
\eta^-({\bf p},{\bf q},{\bf s})
& \equiv &
\prod_s \eta^-_s({\bf p}^{N_s},{\bf q}^{N_s})
\nonumber \\ & = &
\prod_s
\sum_{\hat{\mathrm P}_s} (- 1)^{p_s}
e^{-[ {\bf p}^{N_s}-{\bf p}'^{N_s}] \cdot {\bf q}^{N_s}/\mathrm{i}\hbar} .
\end{eqnarray}
Because only permutations between same spin fermions contribute,
one can factorize the permutation operator,
and hence also the symmetrization function.

%%%%%%%%%%%%%%%%%%%%%%%%%%%%%%%%%%%%%%%%%%%%%%%%%%%%%%%%%%%%%%%%%%%%
\subsection{Fermion pairs}

\subsubsection{Effective Bosons}

%%%%%%%%%%%%%%%%%%%%%%%%%%%%%%%%%%%%%%%%%%%%%%%%%%%%%%%%%%%%%%%%%%
\begin{figure}[t!]
\centerline{ \resizebox{8cm}{!}{ \includegraphics*{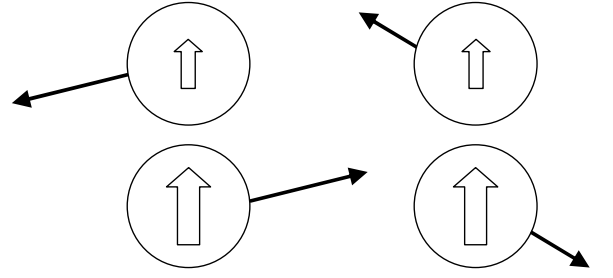} } }
% from My Documents\Projects\QSM22\paired.doc
\caption{\label{Fig:pair}
Paired fermions 1 and 2 (left), and 3 and 4 (right).
%The spatial locations are not relevant to the pairing.
}
\end{figure}
%%%%%%%%%%%%%%%%%%%%%%%%%%%%%%%%%%%%%%%%%%%%%%%%%%%%%%%%%%%%%%%%%%

The  usual definition of a Cooper pair is that
the momenta must be equal and opposite,
${\bf p}_1 = -{\bf p}_2$, and also the spins, $s_1 = - s_2$.\cite{Cooper56}
This is widely accepted to act as a zero momentum, zero spin effective boson.
The more general definition that will be used here
likewise insists that the two fermions have  equal and opposite momenta,
${\bf p}_1 = -{\bf p}_2$.
But for the the spins it is only necessary that $s_1 \ne s_2$
in order to prevent internal permutations within the Cooper pair,
which would cancel the bosonic ones.
For the case of electrons, which are spin-half fermions,
the condition  $s_1 \ne s_2$ is equivalent to  $s_1 = -s_2$.
Electrons of course are the main focus in superconductivity,
in which case there is no difference in the two formulations
of Cooper pairs.
The more general pair may be called an effective spin-$s_1s_2$ boson,
with the convention $s_1 < s_2$.
The total spin $s_1+s_2$ is not sufficient to label the pair.
There are $S(2S+1)$ distinct species of pairs.

Consider four fermions in two  pairs: $\{1,2\}$ and $\{3,4\}$
(see Fig.~\ref{Fig:pair}).
That is, ${\bf p}_1 = -{\bf p}_2$
and ${\bf p}_3 = -{\bf p}_4$.
Suppose that $s_1=s_3$ and $s_2=s_4$.
Then the only permutations permitted are between
1 and 3 and between 2 and 4.
Hence the symmetrization function\cite{Attard18,Attard21}
for these four fermions is
\begin{eqnarray}
\lefteqn{
\sum_{\hat{\mathrm P}}
(-1)^p
e^{-{\bf q} \cdot [{\bf p}-{\bf p}']/\mathrm{i}\hbar}
\delta_{{\bf s}',{\bf s}}
} \nonumber \\
& = &
1
- e^{- {\bf q}_{13} \cdot {\bf p}_{13} /\mathrm{i}\hbar}
- e^{- {\bf q}_{24} \cdot {\bf p}_{24} /\mathrm{i}\hbar}
\nonumber \\ && \mbox{ }
+ e^{- {\bf q}_{13} \cdot {\bf p}_{13} /\mathrm{i}\hbar}
e^{- {\bf q}_{24} \cdot {\bf p}_{24} /\mathrm{i}\hbar}
\nonumber \\ & \approx &
1
%+ e^{- ({\bf q}_{12}-{\bf q}_{34})
%\cdot ({\bf p}_{12}-{\bf p}_{34})  /2\mathrm{i}\hbar}
+
e^{ - {\bf q}_{12} \cdot {\bf p}_{13}/\mathrm{i}\hbar}
e^{ - {\bf q}_{34} \cdot {\bf p}_{31}/\mathrm{i}\hbar}.
\end{eqnarray}
The prime indicates the permuted eigenvalues.
The single transposition fermionic terms
have been neglected in the final equality because
of their rapid fluctuation
compared to the retained terms,
which are bosonic.
The double transposition has exponent that in part is
\begin{eqnarray}
\lefteqn{
{\bf q}_{13} \cdot {\bf p}_{13}
+
{\bf q}_{24} \cdot {\bf p}_{24}
} \nonumber \\
& = &
\frac{1}{2} \big[ ({\bf q}_{1}+{\bf q}_{2}) -({\bf q}_{3}+{\bf q}_{4}) \big]
\cdot {\bf p}_{13}
\nonumber \\ && \mbox{ }
+
\frac{1}{2} \big[ ({\bf q}_{1}-{\bf q}_{2}) -({\bf q}_{3}-{\bf q}_{4}) \big]
\cdot {\bf p}_{13}
\nonumber \\ && \mbox{ }
+ \frac{1}{2} \big[ ({\bf q}_{1}+{\bf q}_{2}) -({\bf q}_{3}+{\bf q}_{4}) \big]
\cdot {\bf p}_{24}
\nonumber \\ && \mbox{ }
-
\frac{1}{2} \big[ ({\bf q}_{1}-{\bf q}_{2}) -({\bf q}_{3}-{\bf q}_{4}) \big]
\cdot {\bf p}_{24}
\nonumber \\ & = &
{\bf Q}_{13} \cdot {\bf P}_{13}
+
\frac{1}{2} ({\bf q}_{12}-{\bf q}_{34})
\cdot ({\bf p}_{13}-{\bf p}_{24})
\nonumber \\ & = &
{\bf q}_{12} \cdot {\bf p}_{13}
+
{\bf q}_{34} \cdot {\bf p}_{31}.
\end{eqnarray}
The first two equalities hold in general;
the final equality holds for the Cooper pairs.
The center of mass separation is
${\bf Q}_{13} = {\bf Q}_{1} - {\bf Q}_{3}
= ({\bf q}_{1}+{\bf q}_{2})/2 - ({\bf q}_{3}+{\bf q}_{4})/2 $,
the total momentum difference is
${\bf P}_{13}
= ({\bf p}_{1}+{\bf p}_{2}) - ({\bf p}_{3}+{\bf p}_{4})$,
and the `locations' of the effective bosons  are
${\bf q}_{12} = {\bf q}_{1} -{\bf q}_{2}  $
and ${\bf q}_{34} = {\bf q}_{3} -{\bf q}_{4} $.
For a Cooper pair the total momentum is identically zero,
which gives the final equality.

In its final form the permutation weight
is exactly the dimer symmetrization weight
for two boson molecules located at
${\bf q}_{12}  $ and ${\bf q}_{34} $
with momenta ${\bf p}_1$ and ${\bf p}_3$, respectively.
\cite{Attard18,Attard21}
This is a non-local expression
since it depends only on the size
(i.e.\ internal separation) of the two Cooper pairs.
If there exists an attractive potential
so that the size of a pair is small,
then the fluctuations in this term are also small.
There are infinitely more pairs of fermions
with macroscopic separations $Q_{13}$
then there are with microscopic separations,
and for these the fermionic terms fluctuate infinitely more rapidly than
the bosonic terms.
The former average to zero; the latter average close to unity.
The Cooper pair formulation
removes the macroscopic separation between the center of masses, ${Q}_{13}$.
This is what makes the permutation of Cooper pairs non-local
and creates the analogy with Bose-Einstein condensation and superfluidity.
\cite{Attard21,Attard22a}

%%%%%%%%%%%%%%%%%%%%%%%%%%%%%%%%%%%%%%%%%
\subsubsection{Pair Weight for Bound Cooper Pairs}

One can demonstrate the idea
that a Cooper pair is an effective boson molecule
by performing the classical momentum integral
for the four fermions comprising the above pair dimer.
The  momentum integral is
\begin{eqnarray}
I_4
& \equiv &
\Delta_p^{-6}
\int \mathrm{d}{\bf p}^4 \;
\delta({\bf p}_1+{\bf p}_2)\, \delta({\bf p}_3+{\bf p}_4)\,
 \\ && \mbox{ } \times
e^{-\beta {\cal K}({\bf p}^4)}
e^{-{\bf q}_{12}\cdot{\bf p}_{13}/\mathrm{i}\hbar}
e^{-{\bf q}_{34}\cdot{\bf p}_{31}/\mathrm{i}\hbar}
\nonumber \\ & = &
\Delta_p^{-6}
\int \mathrm{d}{\bf p}_1 \, \mathrm{d}{\bf p}_3 \;
e^{-2 \beta p_1^2/2m} e^{-2 \beta p_3^2/2m}
\nonumber \\ && \mbox{ } \times
e^{-[{\bf q}_{12}-{\bf q}_{34}]\cdot{\bf p}_{1}/\mathrm{i}\hbar}
e^{-[{\bf q}_{34}-{\bf q}_{12}]\cdot{\bf p}_{3}/\mathrm{i}\hbar}
%\nonumber \\ & = &
%V^2 (2\pi\hbar)^{-6} (2\pi m/2\beta)^3
%e^{-(\beta/m) [{\bf q}_{12}-{\bf q}_{34}]^2( m/2\beta \hbar)^2 }
%\nonumber \\ && \mbox{ } \times
%e^{-(\beta/m) [{\bf q}_{34}-{\bf q}_{12}]^2( m/2\beta \hbar)^2 }
\nonumber \\ & = &
\left\{ 2^{-3/2} V \Lambda^{-3}
e^{- \pi [{\bf q}_{12}-{\bf q}_{34}]^2/2\Lambda^2 } \right\}^2
\nonumber \\ & \approx &
2^{-3/2} V \Lambda^{-3}
%\nonumber \\ && \mbox{ } \times
\frac{\Lambda e^{- \pi q_{12}^2/\Lambda^2 }}{ q_{12}\sqrt{2\pi} }
\sinh^{1/2}( 2\pi q_{12}^2 /\Lambda^2 )
\nonumber \\ && \mbox{ } \times
2^{-3/2} V \Lambda^{-3}
\frac{\Lambda e^{- \pi q_{34}^2/\Lambda^2 }}{q_{34} \sqrt{2\pi} }
\sinh^{1/2}( 2\pi q_{34}^2/\Lambda^2 )  .\nonumber
\end{eqnarray}
%The  factor of $2^{-3/2} \Lambda^{-3}V$
%from the momentum integral holds for all Cooper pairs
%and will be included explicitly below.
Since the permutations of Cooper pairs are non-local,
the vast majority are at macroscopic separations
and are therefore uncorrelated.
The final equality
follows by  averaging the coupling exponential over the alignment angle,
\begin{eqnarray}
\lefteqn{
\frac{1}{2}
\int_{-1}^1 \mathrm{d}x \; e^{ 2\pi q_{12}q_{34} x/\Lambda^2 }
} \nonumber \\
& = &
\frac{\Lambda^2}{2\pi q_{12}q_{34}}
\sinh\!\big( \frac{2\pi q_{12}q_{34}}{\Lambda^2} \big)
\nonumber \\ & \approx &
\frac{\Lambda}{ q_{12}\sqrt{2\pi} }
\sinh^{1/2}( 2\pi q_{12}^2 /\Lambda^2 )
\nonumber \\ && \mbox{ } \times
\frac{\Lambda}{q_{34} \sqrt{2\pi} }
\sinh^{1/2}( 2\pi q_{34}^2/\Lambda^2 ) .
\end{eqnarray}
The second equality is valid if $q_{12} \approx q_{34}$,
which is the case when the pairs are bound at the minimum
in the pair potential of mean force $\overline q$.
%The second equality is also valid if the sizes
%of the Cooper pairs are much less
%than the thermal wave length,
%$2\pi q_{12} q_{34}  \ll \Lambda^2$.

In this formulation the weight of the pair dimer
factorizes into the product of weights of each pair.
A mean field approximation
(fix the neighbors in the most likely parallel configuration
and average %the orientation of
the intervening pair successively around the permutation loop)
shows that a similar factorization holds for the pair trimer, etc.
Therefore each Cooper pair
is a boson molecule with average  internal weight
due to the transposition
\begin{eqnarray} \label{Eq:numf}
\nu_\mathrm{mf}
& \equiv &
\left\langle e^{-\pi q_{12}^2/\Lambda^2}
\frac{\Lambda}{q_{12} \sqrt{2\pi} }
\sinh^{1/2}( 2\pi q_{12}^2/\Lambda^2 )  \right\rangle_\mathrm{bnd}
\nonumber \\  & \approx &
\frac{\Lambda e^{-\pi \overline q^2/\Lambda^2} }{\overline q \sqrt{2\pi} }
\sinh^{1/2}( 2\pi \overline q^2/\Lambda^2 ) .
\end{eqnarray}
This holds for a bound Cooper pair
in which the pair potential of mean force
has a relatively narrow minimum at $\overline q$.
This expression for the internal weight
for the bound Cooper pair effective boson is less than unity;
the expression itself is likely an overestimate
due to the mean field approximation (see the Appendix).
The internal weight goes to zero
as the size of Cooper pair goes to infinity.
The remaining factor of $2^{-3/2} \Lambda^{-3}V$
from the momentum integral holds for all Cooper pairs
and will be included explicitly below.

The factorization is valid if the sizes of the bound Cooper pairs
are all at the minimum of the pair potential of mean force
$\overline q$.
This means that the departure from the minimum $\Delta_q$
must be small,
\begin{equation}
\Delta_q \ll \overline q/2
\mbox{ and }
\Delta_q  \overline q \ll \Lambda^2 /2.
\end{equation}
This is the case if the curvature of the pair potential of mean force
is large enough, $\beta \overline w'' \Delta_q ^2/2 \gg 1$.
This is  necessary for the existence of bound Cooper pairs.

%%%%%%%%%%%%%%%%%%%%%%%%%%%%%%%%%%%%%%%%%%%%%%%%%
\subsubsection{Number of Bound Cooper Pairs}

Distinguish between Cooper pairs and bound Cooper pairs,
the latter being separated by less than $\overline q +\Delta_q$.
There are $N_{0,s's''}$ Cooper $s's''$-pairs,
and $N_{0,s's''}^\mathrm{bnd}$ bound Cooper $s's''$-pairs.

The number of bound  fermion couples without regard to spin or momentum is
\begin{equation}
N_\mathrm{bnd}
=
\frac{\rho^2 V}{2}  v_\mathrm{bnd},
\end{equation}
where the bound volume is defined as
\begin{eqnarray} \label{Eq:vbnd1}
v_\mathrm{bnd}
& \equiv &
4\pi
\int_0^{\overline q + \Delta_q} \mathrm {d} q \; q^2 g(q)
\nonumber \\ & \approx &
4\pi \overline q^2
 e^{-\beta \overline w}
\sqrt{2\pi/\beta \overline w''} .
\end{eqnarray}
This assumes that the radial distribution function,
$g(q) = e^{-\beta w(q)}$,
is sufficiently sharply peaked about the minimum
in the pair potential of mean force
to enable a second order expansion
and evaluation of the Gaussian integral.
Also the core exclusion region is assumed close enough
to the minimum that the lower limit can be extended to zero.
It is not essential to make the Gaussian approximation to the integral,
since it can be evaluated numerically by fixing
$\overline q + \Delta_q$ at the barrier to the potential of mean force,
if it has one (see discussion in \S~\ref{Sec:Disco}).
But the Gaussian results do enable a transparent analysis
and a physical interpretation
of the nature of the bound Cooper pairs.

The ratio of Cooper pairs to the number of fermions
must be the same as the ratio of bound Cooper pairs
to the number of bound fermions.
Hence one must have
\begin{equation}
N_{0,s's''}^\mathrm{bnd} =\frac{ N_\mathrm{bnd} }{N} N_{0,s's''} .
\end{equation}
%Each factor here is extensive.
This is an important result.

%%%%%%%%%%%%%%%%%%%%%%%%%%%%%%%%%%%%%%%%%%%%%%%%%
\subsubsection{Symmetrization Function for Cooper Pairs}

The symmetrization function for Cooper pairs is dominated by
the bound Cooper pairs and is
\begin{equation} \label{Eq:eta+0}
\eta^+_{0,s's''}
\approx
\nu^{N_{0,s's''}^\mathrm{bnd}}
N_{0,s's''}^\mathrm{bnd}! .
\end{equation}
The superscript + %for the Cooper pair symmetrization function
indicates that these are treated as effective bosons.
The result follows because the permutations of the bound Cooper pairs
are non-local and each carries the internal weight
of the effective boson, $\nu$.

%%%%%%%%%%%%%%%%%%%%%%%%%%%%%%%%%%%%%%%%%%%%%%%%%
\subsection{Grand Potential with Paired Fermions}

\subsubsection{Continuum Momentum Limit}

In my treatment of superfluidity,\cite{Attard22a}
I discussed taking the continuum momentum limit
by adding the discrete momentum ground state explicitly
to the continuum integral over the supposedly excited momentum states.
Although it appears that the continuum integral
also counts the momentum ground state,
it seems that the double counting has negligible effect
because, at least in the case of ideal bosons,
each case dominates the other in its regime of applicability.\cite{Attard22a}
In any case the formulation that adds
the discrete momentum ground state explicitly
to the continuum integral over excited states
has become well established
ever since it was apparently invoked by London
in his original ideal boson theory of superfluidity.\cite{London38}

One can perform a similar trick for Cooper pairs.
For fermions 1 and 2
write the discrete sum over momentum states as
\begin{eqnarray}
\sum_{{\bf p}_1,{\bf p}_2}
& = &
\sum_{{\bf p}_1}
\left\{ \delta_{{\bf p}_2,-{\bf p}_1}
+ \sum_{{\bf p}_2}\!^{({\bf p}_2\ne -{\bf p}_1)} \right\}
\nonumber \\ & = &
\sum_{{\bf p}_1}
\left\{ \delta_{{\bf p}_2,-{\bf p}_1}
+ \Delta_p^{-3} \int \mathrm{d}{\bf p}_2  \right\}
\nonumber \\ & = &
\frac{1}{\Delta_p^{6}} \int \mathrm{d}{\bf p}_1\, \mathrm{d}{\bf p}_2
\left\{  \Delta_p^{3} \delta({\bf p}_2+{\bf p}_1) + 1   \right\} .
\end{eqnarray}
In transforming to the continuum integral,
I assume that the point
${\bf p}_2 = -{\bf p}_1$ is a set of measure zero
and so this formulation it is not really double counting.
In this form the integral covers the possible states of the two fermions
as both paired and unpaired.
A similar procedure for the product of momentum sums for $N$ fermions
leads to a binomial expansion,
the terms of which consist of a certain number of paired
and a certain number of unpaired fermions,
as is  now derived.

%%%%%%%%%%%%%%%%%%%%%%%%%%%%%%%%%%%%%%%%%%%%%%%%%%%%%%%
\subsubsection{Numbers of Paired and Unpaired Fermions}

Let $N_s$ be the total number of spin-$s$ fermions,
and let $N_{1,s}$ be the number of unpaired  spin-$s$ fermions.
Let $N_{0,ss'}$ be the number of paired fermions
with spin pair $ss'$.
To count pairs uniquely,  $s < s'$;
for convenience I define $ N_{0,ss'} = N_{0,s's} $
and $N_{0,ss} = 0$.
Obviously
\begin{equation}
N_s
=
N_{1,s}
+ \sum_{s'=-S}^S  N_{0,ss'} .
\end{equation}
Also $N = \sum_{s=-S}^S N_s$.

In the first instance take the Hamiltonian to be independent of spin,
in which case $\overline N_s = N/(2S+1)$.
This requirement is relaxed when a spin-dependent potential
is included.

I shall treat the paired and unpaired fermions
as $(S+1)(2S+1)$ different species,
and only allow permutations within each species.
This is is similar to the no mixing approximation
that I have used in the treatment of superfluidity.\cite{Attard22a}
For the occupancy
$\{ \underline N_0 , \underline N_1 \}
\equiv \{N_{0,ss'},N_{1,s}\}$
the total number of permutations restricted to each species is
\begin{eqnarray}
M
& = &
\prod_{s=-S}^S  N_{1,s}!  \prod_{s'=-S}^S \prod_{s''=s'+1}^S N_{0,s's''}!
\nonumber \\  & = &
\prod_{s,s'<s''}  N_{1,s}! \, N_{0,s's''}!.
\end{eqnarray}
The second equality defines the abbreviated notation that will be used.
This permutation number $M$ goes directly into the denominator
in conjunction with the symmetrization factor formalism
for the partition function that treats the paired and unpaired fermions
as different species.

One can also see this from the binomial expansion
that arises from the transformation to the continuum mentioned above.
The usual multinomial factor that arises in the expansion is
\begin{equation}
\prod_{s,s'<s''}
\frac{N_{s}! }{N_{1,s}! \,N_{0,s's''}!  } .
\end{equation}
Inserting this into the grand partition function, Eq.~(\ref{Eq:Xi-}),
the numerator here cancels with the denominator there,
leaving $M$ as the denominator in that equation.

%%%%%%%%%%%%%%%%%%%%%%%%%%%%%%%%%%%%%%%%%%%%%%%%%%%%%%%
\subsubsection{Grand Partition Function}

In view of either of these two results, the grand partition function
for paired and unpaired fermions  is
\begin{eqnarray}
%\lefteqn{
\Xi^-
%} \nonumber \\
 & = &
\sum_{\underline N_{0},\underline N_{1}}
\frac{z^N}{ \prod_{s,s'<s''}  N_{1,s}! \, N_{0,s's''}! }
\nonumber \\ && \mbox{ } \times
\sum_{\bf p}
%\prod_{s'<s''} \sum_{\hat{\mathrm P}_{0,s's''}}
%\prod_{s} \sum_{\hat{\mathrm  P}_{1,s}}
\sum_{\hat{\mathrm  P}}
(-1)^{p_{1,s}}
%\nonumber \\ && \mbox{ } \times
\left \langle \phi_{\bf p}({\bf q})
\left| e^{-\beta \hat {\cal H} } \right|
\phi_{\bf p}\big(\hat{\mathrm P}{\bf q}\big) \right\rangle
\nonumber \\ & \approx &
\sum_{\underline N_{0},\underline N_{1}}
\frac{z^NV^{-N}}{\prod_{s,s'<s''}  N_{1,s}! \, N_{0,s's''}!}
\nonumber \\ && \mbox{ } \times
\sum_{{\bf p}} \int \mathrm{d}{\bf q}\;
e^{-\beta {\cal H}({\bf q},{\bf p})}
\eta^-({\bf q},{\bf p},{\bf s}).
\end{eqnarray}
The commutation function has been neglected in the final equality.
The discrete momentum sums can be replaced directly
by continuum momentum integrals
since the paired and unpaired fermions have been explicitly identified.
The phase space point of the $ss'$ paired fermions
is denoted ${\bf \Gamma}^{2N_{0,ss'}} \equiv
 \{ {\bf q}^{2N_{0,ss'}},{\bf p}^{2N_{0,ss'}} \}$,
and the phase space point of unpaired $s$ fermions is denoted
${\bf \Gamma}^{N_{1,s}} \equiv \{ {\bf q}^{N_{1,s}},{\bf p}^{N_{1,s}} \}$.
The total permutator is
$ \hat{\mathrm P} =
\prod_{s,s'<s''} \hat{\mathrm P}_{0,s's''} \hat{\mathrm P}_{1,s} $,
which factors commute.

Because permutations are confined to amongst fermions with the same spin,
and because of the no mixing approximation
(i.e.\ permutations between paired and unpaired fermions may be neglected),
the symmetrization function fully factorizes
\begin{equation}
\eta^-({\bf q},{\bf p},{\bf s})
=
\prod_{s,s'<s''}  \eta^+_{0,s's''}({\bf \Gamma}^{2N_{0,s's''}}) \,
\eta^-_{1,s}({\bf \Gamma}^{N_{1,s}}).
\end{equation}
The  symmetrization function for paired fermions, $\eta_{0,s's''}^+$,
was expressed above in terms of the number of bound Cooper pairs,
Eq.~(\ref{Eq:eta+0}).
It is independent of the point in phase space
and can be taken outside of the integrals for the partition function.

The  symmetrization function for unpaired fermions is
\begin{equation}
\eta^-_{1,s}({\bf \Gamma}^{N_{1,s}})
=
\sum_{\hat{\mathrm P}_{1,s}}^{N_{1,s}!} (-1)^{p_{1,s}}
e^{-[ {\bf p}^{N_{1,s}}-{\bf p}'^{N_{1,s}}]
\cdot {\bf q}^{N_{1,s}} /\mathrm{i}\hbar} .
\end{equation}
Here ${\bf p}'^{N_{1,s}} \equiv \hat{\mathrm P}_{1,s}{\bf p}^{N_{1,s}}$,
and $p_{1,s}$ is the parity of the permutation $\hat{\mathrm P}_{1,s}$.

In the general classical phase space formulation
of quantum statistical mechanics,
the permutation loop expansion,
which consists of products of loops,
can be written as the exponential of a series of single loops.
\cite{Attard18,Attard21}
It follows that the grand potential is the sum of loop potentials,
each of which is the classical average of a sum over single permutation loops
of a given size.
This holds only for the unpaired fermions,
whose symmetrization function has $(2S+1)$ factors.
Each factor
gives a corresponding loop grand potential;
for $l \ge 2$ these can all be treated independently.

%%%%%%%%%%%%%%%%%%%%%%%%%%%%%%%%%%%%%%%%%%%%%%%%%%%%%%%%%%%%%%%%%%%%
\subsubsection{Configuration Integral Containing Bound Cooper Pairs}

Shortly the monomer or classical grand potential will be given,
and this depends upon the classical  configurational integral
containing $N^\mathrm{bnd}_{0} = \sum_{s'<s''} N^\mathrm{bnd}_{0,s's''}$
bound Cooper pairs.
This is different to the classical  configurational integral $Q(N,V,T)$
unconstrained by such pairing.
The two are related as
\begin{eqnarray} \label{Eq:vbnd}
\lefteqn{
Q(N,V,T|\underline N^\mathrm{bnd}_0)
} \nonumber \\
& = &
\int_V \mathrm{d}{\bf q}^{N_1} \mathrm{d}{\bf q}^{2N_0-N_0^\mathrm{bnd}}
%\nonumber \\ && \mbox{ } \times
\int_{\Delta_V}  \mathrm{d}{\bf q}^{N_0^\mathrm{bnd}}\;
e^{-\beta U({\bf q}^N)}
\nonumber \\ & = &
Q(N,V,T)
\left\langle \prod_{j=1}^{N_0^\mathrm{bnd}}
\Theta( \overline q + \Delta_q  - q_{jj'})
\right\rangle
\nonumber \\ & = &
Q(N,V,T) \left\{ \frac{1}{V} 4\pi
\int_0^{ \overline q + \Delta_q } \mathrm{d} q\; q^2
g(q)
\right\}^{N_0^\mathrm{bnd}}
%\nonumber \\ & \approx &
%Q(N,V,T)  \left\{ \frac{4\pi}{V}  \overline q^2 e^{-\beta \overline w}
% \sqrt{ 2\pi/\beta \overline w'' } \right\}^{N_0^\mathrm{bnd}}
\nonumber \\ & \equiv &
Q(N,V,T)
\prod_{s'<s''}
\frac{v_\mathrm{bnd}^{N^\mathrm{bnd}_{0,s's''}}
}{V^{N^\mathrm{bnd}_{0,s's''}} } .
\end{eqnarray}
Here $j'$ is the label of the fermion bound to $j$,
and $\Theta(q)$ is the Heaviside step function.
The bound weighted volume $v_\mathrm{bnd}$ was given in Eq.~(\ref{Eq:vbnd1}),
where its proportionality to
the number of bound generic fermions was given,
$N_\mathrm{bnd} = (\rho^2 V /2) v_\mathrm{bnd}$.
%Also, the bound volume can be evaluated numerically
%instead of by the Gaussian approximation.

%%%%%%%%%%%%%%%%%%%%%%%%%%%%%%%%%%%%%%%%%%
\subsubsection{Monomer Grand Potential}

Finally,
the monomer or classical grand potential is
\begin{eqnarray}
\lefteqn{
e^{ -\beta \Omega^{(1)}(z,V,T) }
} \nonumber \\
& = &
\Xi_\mathrm{cl}(z,V,T)
\nonumber \\ & = &
\sum_{\underline N_{0},\underline N_{1}}
\frac{z^N
\prod_{s'<s''}  \nu^{N^\mathrm{bnd}_{0,s's''}} N^\mathrm{bnd}_{0,s's''}!
}{ V^{N} \prod_{s,s'<s''}  N_{1,s}! N_{0,s's''}! }
\nonumber \\ && \mbox{ } \times
\prod_{s,s'<s''}
\frac{\Delta_p^{-3N_{1,s}}}{\Delta_p^{3N_{0,s's''}}}
\int \mathrm{d}{\bf p}^{N_{0,s's''}}\,\mathrm{d}{\bf p}^{N_{1,s}}
e^{ -\beta {\cal K}({\bf p}^{N}) }
\nonumber \\ && \mbox{ } \times
\int_V \mathrm{d}{\bf q}^{N_1} \mathrm{d}{\bf q}^{N_0-N_0^\mathrm{bnd}}
%\nonumber \\ && \mbox{ } \times
\int_{\Delta_V}  \mathrm{d}{\bf q}^{N_0^\mathrm{bnd}}\;
e^{-\beta U({\bf q}^N)}
\nonumber \\ & = &
\sum_{\underline N_{0},\underline N_{1}}
\frac{z^N
\prod_{s'<s''} \nu^{N^\mathrm{bnd}_{0,s's''}} N^\mathrm{bnd}_{0,s's''}!
}{  V^{N} \prod_{s,s'<s''}  N_{1,s}! N_{0,s's''}! }
\nonumber \\ && \mbox{ } \times
\prod_{s,s'<s''}
\frac{ (V/\Lambda^3)^{N_{1,s}}  }{(2^{3/2} \Lambda^{3} /V )^{N_{0,s's''}} } \,
Q(N,V,T|\underline N^\mathrm{bnd}_0)
\nonumber \\ & = &
\sum_{\underline N_{0},\underline N_{1}}
\prod_{s'<s''}
\frac{ (\nu v_\mathrm{bnd})^{N^\mathrm{bnd}_{0,s's''}}
N^\mathrm{bnd}_{0,s's''}!
}{ V^{N^\mathrm{bnd}_{0,s's''}} }
\nonumber \\ && \mbox{ } \times
\prod_{s'<s''}
\frac{ V ^{N_{0,s's''}}
}{ (2^{3/2} \Lambda^{3} )^{N_{0,s's''}} N_{0,s's''}! }
\nonumber \\ && \mbox{ } \times
\prod_{s}
\frac{ (V/\Lambda^3)^{N_{1,s}} }{ N_{1,s}!}
\times
\frac{z^N}{V^N} Q(N,V,T) .
\end{eqnarray}
Here the continuum momentum limit has been taken.
The  classical configuration integral, $Q(N,V,T)$,
is independent of the spin state of the fermions
(unless a spin-dependent potential is added).
As above, $N=\sum_s  N_{1,s} + \sum_{s'<s''}  N_{0,s's''} $.

%%%%%%%%%%%%%%%%%%%%%%%%%%%%%%%%%%%%%%%%%%
\subsubsection{Unpaired Grand Potential}

For unpaired fermions the loop grand potential
is straightforward to derive as, apart from a factor of $(-1)^{l-1}$,
it is identical to that for bosons.
\cite{Attard18,Attard21}
The unpaired loop grand potentials $l \ge 2$ are classical averages
(\S 5.3 of Ref.~\onlinecite{Attard21}),
which can be taken canonically
\begin{eqnarray}
-\beta \Omega^{-,(l)}_{1,s}
& = &
\left\langle \eta^{-,(l)}_{1,s}
({\bf p}^{N_s},{\bf q}^{N_s})
\right\rangle_{\underline N_{0},\underline N_{1}}^\mathrm{cl}
\nonumber \\ & = &
(-1)^{l-1}
\left\langle G^{(l)}({\bf q}^{N_{1,s}})
\right\rangle_{\underline N_{0},\underline N_{1}}^\mathrm{cl}
\nonumber \\ & = &
(-1)^{l-1} \left(\frac{N_{1,s}}{N}\right)^{l}
\left\langle G^{(l)}({\bf q}^{N}) \right\rangle_{N}^\mathrm{cl}
\nonumber \\ & \equiv &
N_{1,s} (-1)^{l-1} \left(\frac{N_{1,s}}{N}\right)^{l-1} g^{(l)}  .
\end{eqnarray}
The fermionic anti-symmetrization factor
$(-1)^{l-1}$ appears explicitly.
Apart from it, all factors are positive.
The average has been transformed  from the mixed
$\{\underline N_{0},\underline N_{1}\}$ system
to the classical configurational system of $N$ fermions
that does not distinguish their state.\cite{Attard21}
The factor of  $\left({N_{s,1}}/{N}\right)^{l}$
 is the uncorrelated probability that $l$ fermions
chosen at random in the original mixed system are all unpaired
and in the same state.
The Gaussian position loop function is
\begin{equation}
G^{(l)}({\bf q}^{N})
=
%\sum_{j_1,j_2,\ldots,j_l}^{N_*} \hspace{-4mm} ' \hspace{3mm}
%e^{-\pi {\cal L}(j_1,j_2,\ldots, j_l)^2 /\Lambda^2 } .
\sum_{j_1,\ldots,j_l}^N\hspace{-.2cm}''\hspace{.1cm}
e^{-\pi q_{j_l,j_1}^2 /\Lambda^2 }
\prod_{k=1}^{l-1}
e^{-\pi q_{j_k,j_{k+1}}^2 /\Lambda^2 }  .
\end{equation}
The double prime indicates that no two indeces may be equal
and that distinct loops must be counted once only.
There are $N!/(N-l)!l$ distinct $l$-loops here,
the overwhelming number of which are negligible upon averaging.
Since the pure excited momentum state permutation loops
are compact in configuration space,
one can define an intensive form
of the average loop Gaussian,
$g^{(l)} \equiv
\left\langle G^{(l)}({\bf q}^{N}) \right\rangle_{N}^\mathrm{cl} /N$.
This is convenient because it does not depend upon $N_{1,s}$.

%%%%%%%%%%%%%%%%%%%%%%%%%%%%%%%%%%%%%%%%%%%%%%%%%%%%%%%%%%%%%%%%%%%%%%%%
\subsection{Maximum Entropy for Paired Fermions}

The constrained grand potential is given by
\begin{eqnarray}
\lefteqn{
-\beta \Omega^{-}(\underline N_0, \underline N_1 |z,V,T)
}  \\ \nonumber
& = &
-\beta \Omega^{(1)}(\underline N_0, \underline N_1 |z,V,T)
-\beta \sum_{s=-S}^S \sum_{l=2}^\infty
\Omega^{-,(l)}_{1,s}(\underline N_0, \underline N_1 ) .
\end{eqnarray}
The monomer and paired
loop potentials on the right hand side are given above.

This is the constrained total  entropy
and the optimum number of paired and unpaired fermions
is determined by maximizing it.
It is most convenient to take the derivatives at constant number $N$.

In the absence of a spin-dependent potential,
all spins are equal, $N_s = N/(2S+1)$.
Similarly the pair number is independent of the spin pair $s's''$,
$N_{0,s's''} = [1-\delta_{s',s''}] \tilde N_0 $, so that
$N_s = N_{1,s} + \sum_{s'} N_{0,ss'}
= N_{1,s} + 2S \tilde N_0 $.
One can ensure the derivative at constant  $N$
by taking
$\mathrm{d} N_{1,s'}
= \mathrm{d} N_{1,s''} = -\mathrm{d} N_{0,s's''}$.
Setting $N_{1,s'} = N_{1,s''}$
after the differentiation one has
\begin{eqnarray} \label{Eq:dWdN}
%\lefteqn{
\frac{\mathrm{d} (-\beta \Omega) }{\mathrm{d}  N_{1,s'}}
% } \nonumber \\
& = &
\ln \frac{ V  N_{0,s's''}}{2^{-3/2} \Lambda^{3}  N_{1,s'}^2}
-
\frac{N_\mathrm{bnd}}{N}
\ln \frac{\nu v_\mathrm{bnd} N_\mathrm{bnd}  N_{0,s's''}}{NV}
\nonumber \\ && \mbox{ }
+ 2 \sum_{l=2}^\infty (-1)^{l-1} l
 \frac{N_{1,s'}^{l-1}}{N^{l-1}}  g^{(l)}  .
\end{eqnarray}
Substituting  $N_{0,s's''} = (N_{s'} - N_{1,s'} )/2S $
and setting the derivative to zero
determines $\overline N_{1,s'}$
for fixed $N$ and $V$.
If the right hand side is positive,
then  $N_{1,s'}$ should be increased.

At high temperatures,
the intensive Gaussian loop integrals are negligible,
$g^{(l)} \approx 0$,
as are the number of bound generic pairs, $N_\mathrm{bnd}$.
In this regime
the solution corresponds to the argument of the first logarithm being unity,
or
\begin{equation}
\overline N_{0,s's''}
= 2^{-3/2}  \Lambda^{3} \overline \rho_{1,s'}\,
\overline N_{1,s'} .
\end{equation}
One sees that the number of Cooper pairs is relatively negligible
until the temperature is low enough that
$  \rho  \Lambda^3 \agt 2^{3/2}  $
(assuming that the intensive Gaussian loop integrals remain negligible).

%%%%%%%%%%%%%%%%%%%%%%%%%%%%%%%%%%%%%%%%%%%%%%%%%%%%%%%%%%%%%%%%%%%%%%%%
\subsubsection{Spin Dependent External Potential}

Add an external spin-dependent one-body potential,
$U(\sigma) = -B\sum_j \sigma_j$.
Assume that there is no pair or many-body potential
that depends upon the spins.
The classical configuration integral becomes
\begin{eqnarray}
%\lefteqn{
Q(N,V,T,B)
%} \nonumber \\
& = &
\prod_{s,s'<s''} e^{ s \beta B N_{1,s} }
e^{  (s'+s'') \beta B  N_{0,s's''}  }
\nonumber \\ & & \mbox{ } \times
Q(N,V,T).
\end{eqnarray}
To the monomer grand potential $-\beta \Omega^{(1)}$
should be added
%constrained monomer grand potential (actually $-\beta$ times this)
\begin{equation}
\beta B \sum_{s}  s N_{1,s}
+ \beta B \sum_{s'<s''}  (s'+s'')  N_{0,s's''} .
\end{equation}

In general the derivatives are messy because the
$N_{1,s}$ and $N_{0,s's''}$ vary with spin.
But for electrons, $S=1/2$, one can carry out the derivatives
at constant  $N$ in two ways to give two equations for two unknowns.
The first way is to take
$\mathrm{d} N_{1,\uparrow}
= \mathrm{d} N_{1,\downarrow} = -\mathrm{d} N_{0,\uparrow\downarrow}$.
This gives
\begin{eqnarray}
%\lefteqn{
\frac{\mathrm{d} (-\beta \Omega) }{\mathrm{d}  N_{1,\uparrow}}
% } \nonumber \\
& = &
\ln \frac{ V }{\Lambda^{3}  N_{1,\uparrow}}
+
\ln \frac{ V }{\Lambda^{3}  N_{1,\downarrow}}
-
\ln \frac{ 2^{-3/2} \Lambda^{-3}  V }{N_{0,\uparrow \downarrow}}
\nonumber \\ && \mbox{ }
-
\frac{N_\mathrm{bnd}}{N}
\ln \frac{\nu v_\mathrm{bnd} N_\mathrm{bnd}  N_{0,\uparrow \downarrow}}{NV}
\nonumber \\ && \mbox{ }
+  \sum_{l=2}^\infty (-1)^{l-1} l
 \frac{N_{1,\uparrow}^{l-1}+N_{1,\downarrow}^{l-1}}{N^{l-1}}  g^{(l)} .
%\nonumber \\ && \mbox{ }
%+ \beta B ( s_\uparrow+s_\downarrow)
%- \beta B  (s_\uparrow+s_\downarrow)  .
\end{eqnarray}
The second way is to take
$\mathrm{d} N_{1,\uparrow}
= -\mathrm{d} N_{1,\downarrow}$
and $\mathrm{d} N_{0,\uparrow \downarrow} = 0$,
which gives
\begin{eqnarray}
%\lefteqn{
\frac{\mathrm{d} (-\beta \Omega) }{\mathrm{d}  N_{1,\uparrow}}
% } \nonumber \\
& = &
\ln \frac{ V }{\Lambda^{3}  N_{1,\uparrow}}
-
\ln \frac{ V }{\Lambda^{3}  N_{1,\downarrow}}
+ \beta B
 \\ && \mbox{ }\nonumber
+  \sum_{l=2}^\infty (-1)^{l-1} l
 \frac{N_{1,\uparrow}^{l-1}-N_{1,\downarrow}^{l-1}}{N^{l-1}}  g^{(l)} .
%\nonumber \\ && \mbox{ }
\end{eqnarray}
The number of paired electrons can be eliminated from these by writing
$ N_{0,\uparrow\downarrow} = [ N - N_{1,\uparrow} - N_{1,\downarrow}]/2$.

%%%%%%%%%%%%%%%%%%%%%%%%%%%%%%%%%%%%%%%%%%%%%%%%%%%%%%%%%%%%%%%%%%%%%%%%%%
%
\section{Results}
\setcounter{equation}{0} \setcounter{subsubsection}{0}
%\renewcommand{\theequation}{\arabic{section}.\arabic{equation}}
%\renewcommand{\theequation}{\Alph{section}.\arabic{equation}}
%
%%%%%%%%%%%%%%%%%%%%%%%%%%%%%%%%%%%%%%%%%%%%%%%%%%%%%%%%%%%%%%%%%%%%%%%%%%

% helium-3 has an overall spin of one half
%The transition to a superfluid occurs at 2.491 millikelvins
%on the melting curve.
%In a zero magnetic field, there are two distinct superfluid phases of 3He,
%the A-phase and the B-phase.

%%%%%%%%%%%%%%%%%%%%%%%%%%%%%%%%%%%%%%%%%%%%%%%%%%%%%%%%%%%%%%%%%%
\begin{table}[tb]
\caption{ \label{Tab:gl}
Intensive loop Gaussian for saturated liquid Lennard-Jones He$^3$
at various temperatures,
$T^* \equiv k_\mathrm{B} T /\varepsilon$.
}
\begin{center}
\begin{tabular}{c c c c c }
\hline\noalign{\smallskip}
$ T^* $ & $g^{(2)}$ & $g^{(3)}$  & $g^{(4)}$  & $g^{(5)}$ \\
\hline \rule{0cm}{0.4cm}%
1.2 & 5.43E-03 & 2.22E-04 & 1.67E-05 & 1.77E-06 \\
1.1 & 1.11E-02 & 7.84E-04 & 9.69E-05 & 1.60E-05 \\
1.0 & 2.09E-02 & 2.40E-03 & 4.58E-04 & 1.14E-04 \\
0.9 & 3.80E-02 & 6.85E-03 & 1.92E-03 & 6.93E-04 \\
0.8 & 6.99E-02 & 2.00E-02 & 8.47E-03 & 4.64E-03 \\
0.7 & 1.27E-01 & 5.73E-02 & 3.65E-02 & 2.97E-02 \\
0.6 & 2.31E-01 & 1.64E-01 & 1.59E-01 & 1.94E-01 \\
\hline
\end{tabular} \\
\end{center}
\end{table}
%%%%%%%%%%%%%%%%%%%%%%%%%%%%%%%%%%%%%%%%%%%%%%%%%%%%%%%%%%%%%%%%%%

Table \ref{Tab:gl} gives the intensive loop Gaussians
for Lennard-Jones helium-3.
The Lennard-Jones parameters model helium,
$ \varepsilon_\mathrm{He} = 10.22 k_\mathrm{B}$\,J
and $\sigma_\mathrm{He} = 0.2556$\,nm.\cite{Sciver12}
The Monte Carlo simulation algorithm used to obtain the results
is detailed in Ch.~5 of Ref.~\onlinecite{Attard21}.
In general the loop Gaussians increase with decreasing temperature.
They also decrease with increasing loop size,
except at the lowest temperature studied.
This suggests that terminating the loop series with $l^\mathrm{max}= 5$
at these temperatures will give reliable results
since each term is weighted by the corresponding power of
$N_{1,s}/N \le 1/(2S+1)$.

%%%%%%%%%%%%%%%%%%%%%%%%%%%%%%%%%%%%%%%%%%%%%%%%%%%%%%%%%%%%%%%%%%
\begin{figure}[t]
\centerline{ \resizebox{8cm}{!}{ \includegraphics*{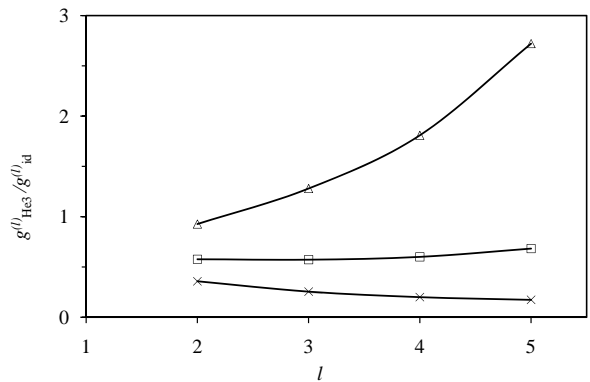} } }
% from My Documents\Projects\QSM22\gloop3He.xlsx
\caption{\label{Fig:gl}
Ratio of intensive loop Gaussian $g^{(l)}$ for Lennard-Jones $^3$He
to that for ideal fermions
at Lennard-Jones helium saturation liquid densities.
The triangles are for $\rho \Lambda^3 = 1.4$ ($T^*=0.6$),
the squares are for $\rho \Lambda^3 = 1.2$ ($T^*=0.7$),
and the crosses are for $\rho \Lambda^3 = 1.1$ ($T^*=0.8$).
Note $T[\mathrm{K}] = 10.22T^*$.
}
\end{figure}
%%%%%%%%%%%%%%%%%%%%%%%%%%%%%%%%%%%%%%%%%%%%%%%%%%%%%%%%%%%%%%%%%%

Figure~\ref{Fig:gl} compares the intensive loop Gaussians $g^{(l)}$
for Lennard-Jones $^3$He and for ideal fermions
by plotting their ratio for  three temperatures.
The intensive loop Gaussian for ideal particles,
 derived in \S 4.1 of Ref.~\onlinecite{Attard21}, is
\begin{eqnarray}
g^{(l)}_\mathrm{id} & = &
\frac{1}{N}
\left\langle G^{(l)}({\bf q}^{N}) \right\rangle_{N,\mathrm{cl,id}}
\nonumber \\ & = &
\frac{1}{N} \frac{N!}{(N-l)!l} \frac{1}{V^l}
\int \, \mathrm{d}{\bf q}^l \;
e^{-\pi q_{l,1}^2 /\Lambda^2 }
\prod_{k=1}^{l-1}
e^{-\pi q_{k,{k+1}}^2 /\Lambda^2 }
%\nonumber \\ & = &
% \frac{N^{l-1}}{lV^{l}}
%\int \, \mathrm{d}{\bf q}^l \;
%e^{-\pi \Lambda^{-2} \underline{\underline G}^{(l)} : {\bf q}^l{\bf q}^l }
%\nonumber \\ & = &
% \frac{\rho^l}{lN}
% \left| (2\pi\Lambda^2/2\pi) \underline{\underline G}^{(l)}  \right|^{-3/2}
% V
\nonumber \\ & = &
\frac{\rho^{l-1}}{l} \Lambda^{3(l-1)} l^{-3/2} .
\end{eqnarray}

It can be seen that at the lowest temperature,
the intensive loop Gaussians the Lennard-Jones model
is larger than that for ideal fermions,
increasingly so as the order of the loop increases.
%This is a graphic example of how
%the particle correlation function
%and potential of mean force
%affects the values of the intensive loop Gaussians,
%particularly at lower temperatures and high densities.
The presence of an attractive potential of mean force
strongly affects the particle correlation functions
and hence the values of the intensive loop Gaussians
that determine the number of Cooper pairs.

%%%%%%%%%%%%%%%%%%%%%%%%%%%%%%%%%%%%%%%%%%%%%%%%%%%%%%%%%%%%%%%%%%
\begin{table}[tb]
\caption{ \label{Tab:param}
Parameters for saturated liquid Lennard-Jones He$^3$ at various temperatures,
$T^* \equiv k_\mathrm{B} T /\varepsilon$.
The Lennard-Jones diameter $\sigma$ is the unit of length.
%Twice the standard error on the mean for the final digit
%is shown in parentheses.
}
\begin{center}
\begin{tabular}{c c c c c c c c}
\hline\noalign{\smallskip}
$ T^* $ & 1.2 & 1.1 & 1.0 & 0.9 & 0.8 & 0.7 & 0.6 \\
\hline \rule{0cm}{0.4cm}%
$\rho$        & 0.53 & 0.63 & 0.70 & 0.75 & 0.80 & 0.85 & 0.89 \\
$\Lambda$     & 1.13 & 1.18 & 1.23 & 1.30 & 1.38 & 1.47 & 1.59 \\
$\rho\Lambda^3$ & 0.59&0.74 & 0.86 & 0.98 & 1.11 & 1.25 & 1.41 \\
$\overline q$ & 1.100 & 1.097 & 1.093 & 1.091 & 1.090 & 1.089 & 1.088 \\
$\beta \overline w $
              &$-0.74$&$-0.81$&$-0.88$&$-0.95$&$-1.03$&$-1.12$&$-1.20$\\
$\beta \overline w'' $
              & 64.49&87.17 &87.39 &100.78&135.60&151.93 &176.07 \\
$\nu$           & 0.32  & 0.33 & 0.35 & 0.37 & 0.39 & 0.42 &  0.45 \\
$v_\mathrm{bnd}$    & 9.92  & 9.12 & 9.68 & 9.63 & 9.03 & 9.25 &  9.38 \\
$\frac{4\pi \overline q}{\Lambda^2\sqrt{\beta \overline w''}} $
              & 1.36  & 1.07 & 0.97 & 0.81 & 0.62 & 0.51 & 0.41 \\

%$\rho \Lambda^3$
%              &0.5912 &0.7430&0.8646&0.9756&1.1065& 1.2489&1.4129 \\
%$g^{(2)}$  &5.43E-03&1.11E-02&2.09E-02&3.80E-02&6.99E-02&1.27E-01&2.31E-01\\
%$g^{(3)}$  &2.22E-04&7.84E-04&2.40E-03&6.85E-03&2.00E-02&5.73E-02&1.64E-01\\
%$g^{(4)}$  &1.67E-05&9.69E-05&4.58E-04&1.92E-03&8.47E-03&3.65E-02&1.59E-01\\
%$g^{(5)}$  &1.77E-06&1.60E-05&1.14E-04&6.93E-04&4.64E-03&2.97E-02&1.94E-01\\
\hline
\end{tabular} \\
\end{center}
\end{table}
%%%%%%%%%%%%%%%%%%%%%%%%%%%%%%%%%%%%%%%%%%%%%%%%%%%%%%%%%%%%%%%%%%

Table~\ref{Tab:param} gives the values of various parameters
obtained by classical canonical simulations of
5,000 Lennard-Jones helium-3 atoms in a homogeneous system
with periodic boundary conditions.
The density is the liquid saturation density,
which was obtained along the saturation curve for a liquid drop.
It can be seen that the minimum in the pair potential of mean force
gets deeper, and its curvature increases, with decreasing temperature.
The internal weight $\nu$ and the bound volume $v_\mathrm{bnd}$
are surprisingly insensitive to temperature.
The parameter
${4\pi \overline q}/{\Lambda^2\sqrt{\beta \overline w''}} $
should be much less than unity for the present numerical approximations
to be valid.
The relation between the Lennard-Jones dimensionless temperature
and the actual temperature is $T[\mathrm{K}] = 10.22T^*$.

%%%%%%%%%%%%%%%%%%%%%%%%%%%%%%%%%%%%%%%%%%%%%%%%%%%%%%%%%%%%%%%%%%
\begin{figure}[t]
\centerline{ \resizebox{8cm}{!}{ \includegraphics*{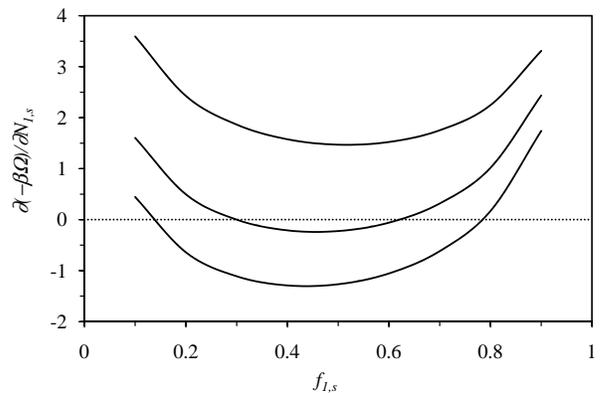} } }
% from My Documents\Projects\QSM22\gloop3He.xlsx
\caption{\label{Fig:F/f1}
Entropy derivative as a function of fraction of unpaired fermions
for Lennard-Jones $^3$He, $S=1/2$,
at various temperatures along the Lennard-Jones saturation liquid curve
at $B=0$.
From bottom to top the curves are $T^*=$ 0.9, 1.0, and 1.1.
The dotted line is a guide to the eye.
Intensive loop Gaussians up to $l^\mathrm{max} = 5$ were used.
%Note $T[\mathrm{K}] = 10.22T^*$.
}
\end{figure}
%%%%%%%%%%%%%%%%%%%%%%%%%%%%%%%%%%%%%%%%%%%%%%%%%%%%%%%%%%%%%%%%%%

Figure \ref{Fig:F/f1} shows the thermodynamic force
acting on the fraction of unpaired fermions,
with positive values driving an increase.
%Again the results are for $^3$He, $S=1/2$.
The curves that don't pass thorough zero
have a positive derivative for the whole domain,
which means that the system in its optimum thermodynamic state
is  composed entirely of unpaired fermions.

For the curves with two zeros,
one is a local maximum and the other is a local minimum in the total entropy.
The stable solution, the local maximum,
is the one with the smaller fraction of unpaired fermions.
Integration of these approximate parabolas shows that the total entropy
is approximately a cubic.
At the highest temperature shown
the total entropy is monotonic increasing with increasing unpaired fraction.
With the emergence of distinct extrema from the point of inflection
slightly above the middle temperature shown, $T^*=1.0$,
the entropy must first decrease for the system
to get from the unpaired state $f_{1,s} = 1$ to
the local entropy maximum at $f_{1,s} \approx 0.3$.
In other words, there is a thermodynamic barrier to the formation
of paired fermions.
This means that the first appearance of a local entropy maximum
at non-zero paired fermion fraction
\emph{cannot} be equated to the condensation transition.
\cite{nb1}

%%%%%%%%%%%%%%%%%%%%%%%%%%%%%%%%%%%%%%%%%%%%%%%%%%%%%%%%%%%%%%%%%%
\begin{figure}[t]
\centerline{ \resizebox{8cm}{!}{ \includegraphics*{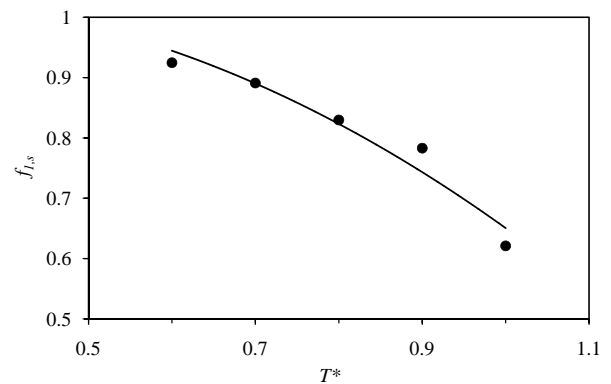} } }
% from My Documents\Projects\QSM22\gloop3He.xlsx
\caption{\label{Fig:olf1}
Fraction of unpaired fermions at the local entropy minimum
for saturated liquid Lennard-Jones $^3$He, $S=1/2$,
as a function of temperature.
The curve is a quadratic best fit constrained
to pass through $f_{1,s} = 1$ at $T^*=0$.
}
\end{figure}
%%%%%%%%%%%%%%%%%%%%%%%%%%%%%%%%%%%%%%%%%%%%%%%%%%%%%%%%%%%%%%%%%%

%The data for the Lennard-Jones fluid, Table~\ref{Tab:param},
%show that it does not fulfil the transition criterion.
The fraction of unpaired $^3$He fermions at the entropy minimum,
Fig.~\ref{Fig:olf1}, increases with decreasing temperature.
The quadratic fit shows that the data
is not incompatible with an entropy minimum
in the fully unpaired system  at $T=0$.
When the entropy minimum reaches $f_{1,s} = 1$,
there is no longer a barrier to the entropy maximum,
and the condensation transition occurs.
The measured condensation transition in $^3$He
occurs at $2.491$\,mK on the melting curve.\cite{Wiki}

%%%%%%%%%%%%%%%%%%%%%%%%%%%%%%%%%%%%%%%%%%%%%%%%%%%%%%%%%%%%%%%%%%
\begin{figure}[t]
\centerline{ \resizebox{8cm}{!}{ \includegraphics*{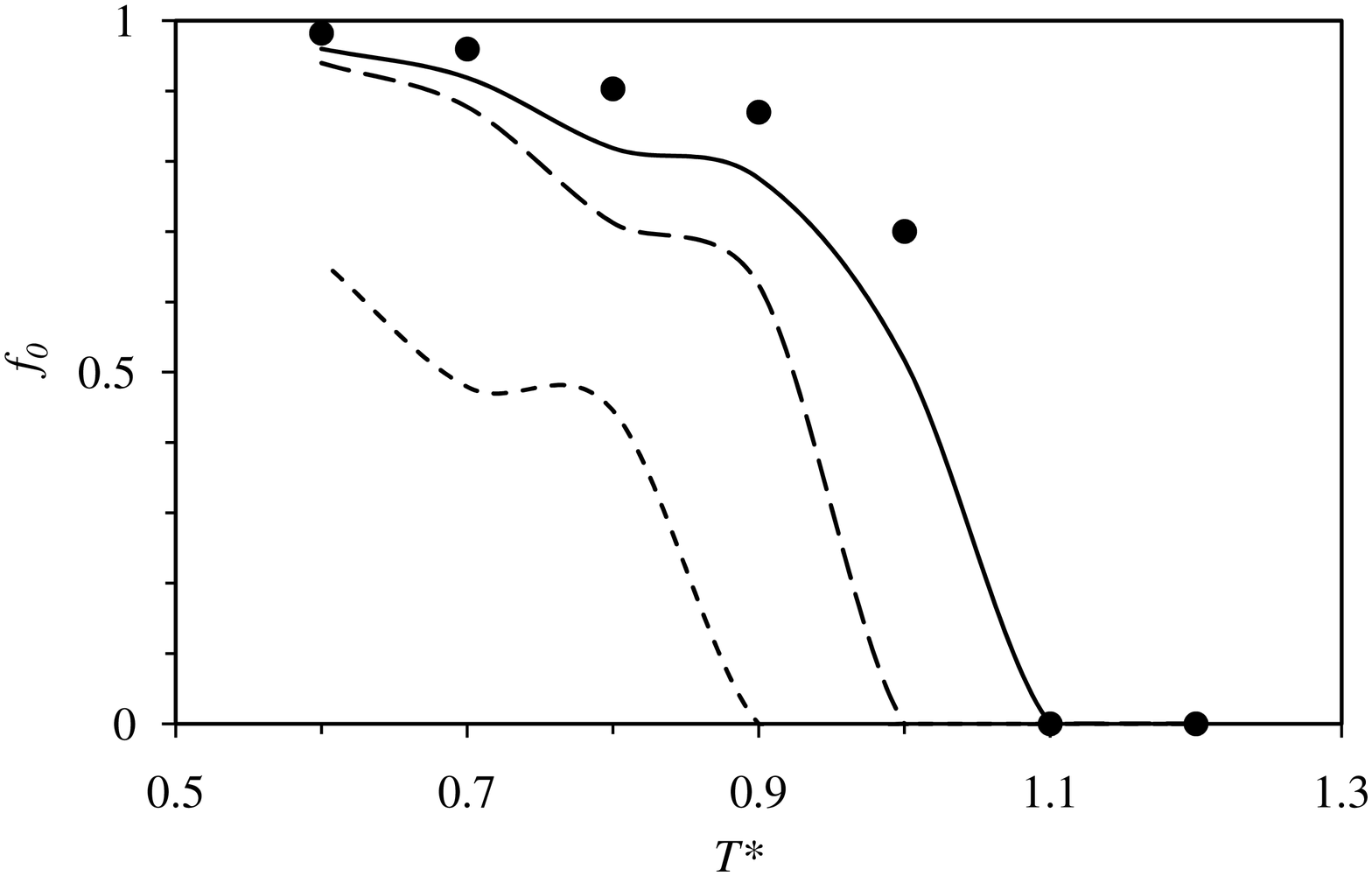} } }
% from My Documents\Projects\QSM22\gloop3He.xlsx
\caption{\label{Fig:olf0}
Fraction of paired fermions at the local entropy maximum
for saturated liquid Lennard-Jones $^3$He, $S=1/2$,
as a function of temperature.
The circles are for $B=0$,
the solid curve is for $\beta B=1$,
the long dashed curve is for $\beta B=2$,
and the short dashed curve is for $\beta B=5$.
%Note $T[\mathrm{K}] = 10.22T^*$.
}
\end{figure}
%%%%%%%%%%%%%%%%%%%%%%%%%%%%%%%%%%%%%%%%%%%%%%%%%%%%%%%%%%%%%%%%%%

Figure \ref{Fig:olf0} shows the fraction of paired $^3$He fermions
at the local entropy maximum as a function of temperature.
It can be seen that in the absence of a spin dependent potential, $B=0$,
this is discontinuous at about $ T^* \approx 1.05$,
or $T = 10.7$\,K.
As just mentioned, this cannot be interpreted as the condensation transition.

Figure \ref{Fig:olf0} also shows the effect
of an external spin-dependent potential,
which can be thought of as an applied magnetic field.
With increasing field strength,
the number of Cooper pairs is reduced,
and the discontinuity temperature is shifted lower.
It is unclear what would occur for $S > 1/2$,
or for an inhomogeneous magnetic field for $S=1/2$.
These tend to align the bound Cooper pairs,
either by the spin dipole or the spin quadrupole,
thereby increasing the internal weight $\nu$,
which would presumably enhance pairing and condensation.

%%%%%%%%%%%%%%%%%%%%%%%%%%%%%%%%%%%%%%%%%%%%%%%%%%%%%%%%%%%%%%%%%%%%%%%%%%
%
\section{Discussion} \label{Sec:Disco}
\setcounter{equation}{0} \setcounter{subsubsection}{0}
%\renewcommand{\theequation}{\arabic{section}.\arabic{equation}}
%\renewcommand{\theequation}{\Alph{section}.\arabic{equation}}
%
%%%%%%%%%%%%%%%%%%%%%%%%%%%%%%%%%%%%%%%%%%%%%%%%%%%%%%%%%%%%%%%%%%%%%%%%%%

Its worth emphasizing that the specific behavior found here
(i.e.\ two zeros in the entropy gradient
and a thermodynamic barrier to condensation)
is very much dependent on the Lennard-Jones potential
and the parameters for $^3$He.
One can isolate the essence of the grand potential derivative
by focussing upon the divergent behavior
at the termini of the domain,
\begin{equation}
\frac{\mathrm{d} (-\beta \Omega) }{\mathrm{d}  N_{1,s}}
\sim
\ln \frac{ (1-f_{1,s})^{1-\rho v_\mathrm{bnd}/2} }{f_{1,s}^2},
\end{equation}
where $f_{1,s}$ is the fraction of unpaired fermions.
The remaining terms are either constant or well behaved loop terms.
As  $f_{1,s} \rightarrow 0$, this is large and positive.
As  $f_{1,s} \rightarrow 1$,
and $ \rho v_\mathrm{bnd} /2 > 1$,
this is also large and positive.
Therefore the derivative has the parabolic shape of Fig.~\ref{Fig:F/f1}.
But if $ \rho v_\mathrm{bnd} /2 < 1$,
then in the limit $f_{1,s} \rightarrow 1$ this is large and negative,
in which case there must be one zero (or an odd number of zeros),
with the one closest to $f_{1,s} = 1$ being an entropy maximum.
In this case there is no thermodynamic barrier to condensation.
Undoubtedly the remaining logarithmic constants and loop grand potentials
shift the zero, but they cannot change this qualitative behavior.
Using the quadratic approximation for the potential of mean force,
as in Eq.~(\ref{Eq:vbnd1}),
the condensation  transition criterion, $\rho v_\mathrm{bnd}/2 < 1$,  is
\begin{equation}
%8\pi^3\rho^2 \overline q^4
%\frac{e^{-2\beta \overline w}}{\beta \overline w''}
%< 1.
2\pi\rho \overline q^2 e^{-\beta \overline w}
\sqrt{\frac{2\pi}{\beta \overline w''} }
< 1.
\end{equation}
One sees that if the pair potential of mean force
acquires a minimum that is short-ranged, shallow, and highly curved,
then the condensation transition will occur.

The data in Table~\ref{Tab:param} show that
a Lennard-Jones fluid cannot satisfy this criterion.
Hence in this case there will always be a thermodynamic barrier
to condensation, albeit one that may shift to
$f_{1,s}=1$ at low temperatures,
at which point it ceases to be a barrier
because the fully unpaired fluid sits at the entropy minimum.

%%%%%%%%%%%%%%%%%%%%%%%%%%%%%%%%%%%%%%%%%%%%%%%%%%%%%%%%%%%%%%%%%%
%\begin{figure}[t]
%\centerline{ \resizebox{8cm}{!}{ \includegraphics*{Fig6.eps} } }
%% from My Documents\Projects\QSM22\paired.doc
%\caption{\label{Fig:pmf}
%Sketch of a pair potential of mean force
%with a barrier to a shallow, narrow primary minimum.
%}
%\end{figure}
%%%%%%%%%%%%%%%%%%%%%%%%%%%%%%%%%%%%%%%%%%%%%%%%%%%%%%%%%%%%%%%%%%

In the BCS theory of superconductivity
the size of the Cooper pairs is hundreds or thousands of times larger than
in the present theory.
It is difficult to see the above criterion,
which scales with $\overline q^2$, being satisfied by BCS theory.
The two theories appear qualitatively different
and likely apply in different regimes.
The fact that BCS theory is a quantum mechanical
rather than a quantum statistical mechanical theory
is consistent with it applying quantitatively
to low temperature superconductors.

%In practical terms the seemingly contradictory requirements
%of a shallow but highly curved minimum in the potential of mean force
%can be satisfied by a repulsive barrier,
%as sketched in Fig.~\ref{Fig:pmf}.
The present quantum statistical theory
likely applies to high temperature superconductors.
The condensation criterion
of a short-ranged, shallow, and highly curved minimum
describes the pair potential of mean force
%for a highly coupled charged system
following the monotonic-oscillatory transition
in a highly coupled charged system
(see Fig.~1 of Ref.~\onlinecite{Attard22b}),
which provides additional evidence
that this transition
%the monotonic-oscillatory transition in a highly coupled charged system
is responsible for high temperature superconductivity.\cite{Attard22b}

The present calculations have several quantitative limitations.
Obviously the Lennard-Jones pair potential is a crude approximation
to the actual interactions in helium.
It is likely inaccurate for low temperature, high density condensed matter
because it neglects the exact short range interactions
and many-body contributions.
Further the present analysis approximated the pair potential of mean force
by a quadratic form and evaluated the Gaussian integrals analytically,
which may not be necessary in a more sophisticated numerical study.
Finally, the present analysis neglected the short-ranged commutation function
on the grounds that Bose-Einstein condensation is dominated
by non-local permutations.
Although this works well for bosons,
the present results for fermions show the essential role
played by the potential of mean force in binding the Cooper pair.
Since the minimum in this occurs at short ranges,
where the function is rapidly varying,
this approximation is questionable.

The present theory reveals the physical basis
and mathematical consequences of Cooper pairing,
and it shows the intimate relationship between
the minimum in the pair potential of mean force and condensation.
The formulation of quantum statistical mechanics
in classical phase space
provides a new approach to
Cooper pair formation, condensation, and superconductivity.

%The analysis shows that a minimum in the pair potential of mean force
%is necessary for the formation of bound Cooper pairs,
%and that bound Cooper pairs are necessary for Bose-Einstein condensation.
The numerical results for Lennard-Jones $^3$He
demonstrate that a minimum in the pair potential of mean force
is not sufficient for condensation.
At higher temperatures there is no entropy maximum for condensed pairs,
and at lower temperatures
there is a thermodynamic barrier
between the local entropy maximum for condensed pairs
and the unpaired state.
The data is consistent with the barrier disappearing close to absolute zero.

More generally the behavior of the constrained total entropy
with fraction of unpaired fermions is very much dependent
on the details of the pair potential of mean force.
A system can go from having a thermodynamic barrier to condensation
to having no barrier  to condensation
if the pair potential of mean force acquires
a sufficiently shallow but highly curved minimum
as the temperature is decreased.
%The most obvious way for this to occur is if there is a barrier
%to the primary minimum in the potential of mean force,
%Fig.~\ref{Fig:pmf}.
This is consistent with %my proposal in Ref.~\onlinecite{Attard22b} that
the monotonic-oscillatory transition in charge systems at high coupling.
\cite{Attard22b}
%and is responsible for the superconducting transition
%in high temperature superconductors.

There are many similarities between superconductivity
and my previous treatment of superfluidity.\cite{Attard22a}
Ultimately both phenomena are a manifestation of Bose-Einstein condensation,
with the paired fermions in the present case being effective bosons.
For bosons,
condensation into the momentum ground state and the consequent superfluidity
is driven by the increase in permutation entropy
due to the non-local correlations
of the momentum ground state.
For paired fermions,
the permutations between the zero-momentum pairs are similarly non-local.
Superfluidity in a bosonic fluid persists because
momentum changing collisions can only occur for the state as a whole,
and hence they have to be of macroscopic, not of molecular, size.
Similarly for superconductivity, because the zero momentum pairs
occupy a single state.
In both cases it is the permutation entropy
that ensures flow without dissipation.

%\newpage
%%%%%%%%%%%%%%%%%%%%%%%%%%%%%%%%%%%%%%%%%%%%%%%%%%%%%%%%%%%%%%%%%%%%%%%%%%
%\section*{References}

%%%%%%%%%%%%%%%%%%%%%%%%%%%%%%%%%%%%%%%%%%%%%%%%%%%%%%%%%%%%%%%%%%%%%%%%%%

%%%%%%%%%%%%%%%%%%%%%%%%%%%%%%%%%%%%%%%

\appendix
%\section{Appendix}

%%%%%%%%%%%%%%%%%%%%%%%%%%%%%%%%%%%%%%%%%%%%%%%%%%%%%%%%%%%%%%%%%%%%%%%%%%
%
\section{Internal Weight}
\setcounter{equation}{0} \setcounter{subsubsection}{0}
\renewcommand{\theequation}{\Alph{section}.\arabic{equation}}
%
%%%%%%%%%%%%%%%%%%%%%%%%%%%%%%%%%%%%%%%%%%%%%%%%%%%%%%%%%%%%%%%%%%%%%%%%%%

%%%%%%%%%%%%%%%%%%%%%%%%%%%%%%%%%%%%%%%%%%%%%%%%%%%%%%%%%%%%%%%%%%
\begin{figure}[b]
\centerline{ \resizebox{8cm}{!}{ \includegraphics*{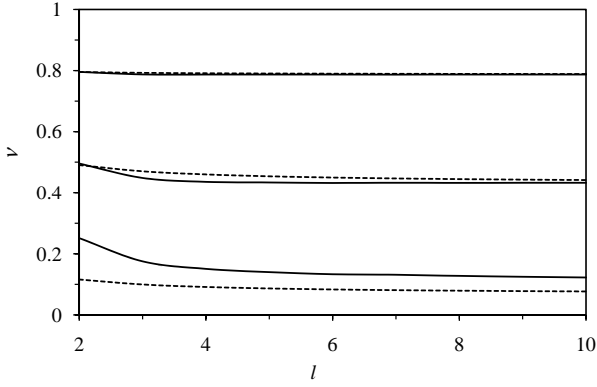} } }
% from My Documents\Projects\QSM22\internal.xlsx
\caption{\label{Fig:nu}
Internal weight per bound Cooper pair of size $\overline q$
as a function of permutation loop length
for $z \equiv \sqrt{\pi}\,\overline q/\Lambda =$
0.5 (top), 1.0 (middle) and 2.0 (bottom).
The solid curves are the exact numerical result,
and the dashed curves are the small $z$ asymptote.
The mean field value is the exact value at $l=2$.
}
\end{figure}
%%%%%%%%%%%%%%%%%%%%%%%%%%%%%%%%%%%%%%%%%%%%%%%%%%%%%%%%%%%%%%%%%%

Figure~\ref{Fig:nu} shows that the exact internal weight
for a bound Cooper pair of size $\overline q$
decreases slowly with increasing loop size from
the mean field result, Eq.~(\ref{Eq:numf}),
which is the exact result at $l=2$.
It appears to reach a plateau for large loop size,
which value would dominate the symmetrization function.
This plateau is confirmed by the exact asymptotic expression,
\begin{equation}
\nu =
e^{-z^2}
\left\{ 1 + \frac{l+2}{6l}  z^4  + {\cal O}(z^8) \right\} ,
\;\;
z \rightarrow 0,
\end{equation}
where $z \equiv \sqrt{\pi}\,\overline q/\Lambda $.
Small $z$ is the relevant regime for Cooper pair condensation
and superconductivity.


\begin{thebibliography}{99}
\bibitem{nb1}
The superfluid transition in $^4$He
shows a similar entropy minimum for small mixed loops,
which thermodynamic barrier disappears for mixed loops with
$l_\mathrm{max} \ge 5$.\cite{Attard22a}
\end{thebibliography}
\end{document}